\documentclass[a4paper,11pt]{article}

\usepackage[english]{babel}
\usepackage[dvips]{graphicx}
\usepackage{amsmath}
\usepackage{amssymb}
\usepackage{epsfig}

\begin{document}

\title{Application of Multifractal Measures to Tehran Price Index}
\author{P. Norouzzadeh $^a$ \footnote{e-mail: noruzzadeh@farda.ir} and G.R. Jafari $^b$
\footnote{e-mail: r.jafari@mehr.sharif.edu}  \\
$^a$Quantitative Analysis Research Group,\\ Farda Development Organization, Tehran, Iran \\
$^b$ Department of Physics, Sharif University of Technology,\\
P.O. Box 11365-9161, Tehran, Iran
\\} \maketitle

\begin{abstract}
We report an empirical study of Tehran Price Index (TEPIX). To
analyze our data we use various methods like as, rescaled range
analysis ($R/S$), modified rescaled range analysis (Lo's method),
Detrended Fluctuation Analysis (DFA) and generalized Hurst
exponents analysis. Based on numerical results, the scaling range
of TEPIX returns is specified, long memory effect or long range
correlation property in this market is investigated,
characteristic exponent for probability distribution function of
TEPIX returns is derived and finally the stage of development in
Tehran Stock Exchange is determined.
\end{abstract}


\section{Introduction}

Financial markets have in recent years been at the focus of
physicists's attempts to apply existing knowledge from statistical
mechanics to economic problems . These markets, though largely
varying in details of trading rules and traded goods, are
characterized by some generic features of their financial time
series, called \emph{stylized facts}. Multifractal processes and
the deeply connected mathematics of large deviations and
multiplicative cascades have been used in many contexts to account
for the time scale dependence of the statistical properties. For
example, recent empirical findings \cite{Stanley, Bouchaud,
FarmerIvory} suggest that in rough surfaces, this framework is
likely to be pertinent. The aim is to characterize the statistical
properties of the series with the hope that a better understanding
of the underlying stochastic dynamics could provide useful
information to create new models able to reproduce experimental
facts. An important aspect concerns concepts as scaling and the
scale invariance of height surface \cite{FarmerIvory, Daco}. There
is an important volume of data and studies showing self-similarity
at short space scales and an apparent breakdown for longer spaces
modeled in terms of distributions with truncated tails. Recent
studies have shown that the traditional approach based on a
Brownian motion picture \cite{Cont, Lux} or other more elaborated
descriptions such as L\'{e}vy and truncated L\'{e}vy distributions
\cite{Bouchaud}, all of them relying on the idea of additive
process, are not suitable to properly describe the statistical
features of these fluctuations. In this sense, there are more and
more evidences that a multiplicative process approach is the
correct way to proceed, and this line of thought leads in a
natural way to multifractality. In fact, this idea was already
suggested some years ago when intermittency phenomena in return
fluctuations was observed at different length scales which gave
rise to some efforts to establish a link with other areas of
physics such as turbulence \cite{Jefferies, Challet}. Nowadays, we
know that there are important differences between both systems, as
for instance the spectrum of frequencies, but the comparison
triggered an intense analysis of the existing data. Multifractal
analysis of a set of data can be performed in two different ways,
analyzing either the statistics or the geometry. A statistical
approach consists of denying an appropriate intensive variable
depending on a resolution parameter, then its statistical moments
are calculated by averaging over an ensemble of realizations and
at random base points. It is said that the variable is
multifractal if those moments exhibit a power-law dependence in
the resolution parameter \cite{Marsili, Giardina}. On the other
hand, geometrical approaches try to assess a local power-law
dependency on the resolution parameter for the same intensive
variables at every particular point (which is a stronger statement
that just requiring some averages—the moments—to follow a power
law). While the geometrical approach is informative about the
spatial localization of self-similar (fractal) structures, it has
been much less used because of the greater technical difficulty to
retrieve the correct scaling exponents. However, in the latest
years an important effort to improve geometrical techniques has
been carried out, giving sensible improvement and good performance
\cite{Farmer, Jefferies-Hart}. We will apply the geometrical
approach in this paper as a valuable tool for the
understanding of the price return fluctuations.\\
The main objective of this article is to investigate the
characteristics of Tehran Stock Exchange using some multifractal
measures. Our purpose is to show how some relatively simple
statistics gives us indications on the market situation. The paper
is organized as follows. In Section 2 we describe our data. In
Section 3, we review the rescaled range ($R/S$) analysis and it's
modified version, Lo's $R/S$ analysis. The way of interpretation
and empirical results for classical $R/S$ analysis and Lo's $R/S$
analysis are presented. In Section 4 a brief description of the
Detrended Fluctuation Analysis is given. The results of this
analysis for the TEPIX time series are shown in this Section too.
In Section 5 we explain the generalized Hurst exponents analysis.
Also its results is presented. In Section 6 the concept of
characteristic exponent for a probability distribution and its
relation to Hurst exponent is reviewed. Moreover, characteristic
exponent of TEPIX returns distribution is computed. Then, in
Section 7, a simple explanation of long memory process is
mentioned and shown how we can use Lo's $R/S$ analysis to find out
presence of such process. The market stage of development and its
relation to generalized Hurst exponent is studied in Section 8.
Finally, conclusions are given in Section 9.

\section{Data Description}
We analyze the values of the TEPIX  for the period of almost 9
years: from 20th may 1995 to 18th march 2004. Before 1995 the
Tehran Price Index was rather fixed because of government
controls. The data have been recorded at each trading day. So that
our database consists of 2342 values and 2341 daily returns. The
sources of this data is the center of research and development of
Iran capital market and the paper utilizes only the closing
prices. In Fig.1 we present a time series corresponding to daily
values of the TEPIX index. It must be mentioned that TEPIX tracks
the performance of more than 350 listed local firms.

\begin{figure}[ht]
{\centering
\resizebox*{0.7\textwidth}{0.35\textheight}{\includegraphics{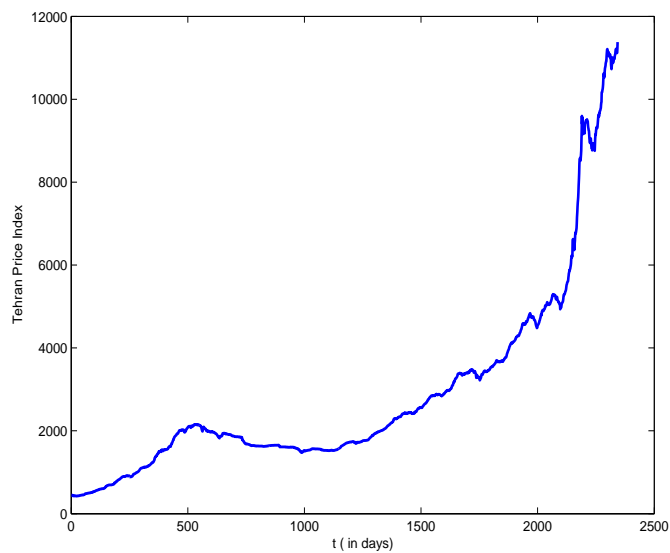}}
\par} \caption{Daily closure TEPIX index history (1995-2004).}
\end{figure}

\vspace*{1cm} Also, Table 1 provides summary statistics of
logarithmic returns.
\begin{table}[htb]
\begin{center}
\caption{\label{Tb2}Mean, standard deviation, skewness, and
kurtosis of TEPIX returns.}
\medskip
\begin{tabular}{ccccccccccccc}
\hline\hline $Mean$&$Std. Dev.$&$Skewness$&$Kurtosis$\\\hline
0.0011 & 0.0046 & 1.0619 & 20.827 \\
\hline\hline
\end{tabular}
\end{center}
\end{table}
According to data in Table 1, it is seen that, the  probability
distribution of TEPIX returns does not look like a Gaussian
distribution and belongs to stable L\'{e}vy distributions.

\section{Rescaled Range ($R/S$) analysis}
\subsection{Classical $R/S$ Analysis}
The Hurst rescaled range ($R/S$) analysis is a technique proposed
by Henry Hurst in 1951 \cite{Hurst} to test presence of
correlations in empirical time series. The main idea behind the
$R/S$ analysis is that one looks at the scaling behavior of the
rescaled cumulative deviations from the mean, or the distance the
system travels as a function of time. This is compared to the
null-hypothesis of a random walk. For an independent system, the
distance covered increases, on average, by the square root of
time. If the system covers a larger distance than this, it cannot
be independent by definition, and the changes must be influencing
each other; they have to be correlated. Although there may be
autoregressive process present that can cause short-term
correlations, we will see that when adjusting for such short-term
correlations, there may be other forms of memory effects present
which need to be examined.\\
Consider a time series in prices of length $P$. This time series
is then transformed into a time series of logarithmic returns of
length $N=P-1$ such that
\begin{equation}
N_{i}=\log(\frac{ P_{i+1} }{P_{i}}), \hspace*{1cm} i = 1, 2, ...,
P - 1.
\end{equation}
Time period is divided into $m$ contiguous sub-periods of length
$n$, such that $m*n = N$. Each sub-period is labelled by $I_{a}$,
with $a=1,2,...m$. Then, each element in $I_{a}$ is labelled by
$N_{k}$, a such that $k = 1, 2, ..., n$. For each sub-period
$I_{a}$ of length $n$ the average is calculated as
\begin{equation}
M_{a}=\frac{1}{n}\sum_{k=1}^{\tau}N_{k,\hspace*{1mm} a}
\end{equation}
Thus, $M_{a}$ is the mean value of the $N_{i}'s$ contained in the
sub-period $I_{a}$ of length $n$. Then, we calculate the time
series of \emph{accumulated departures from the mean}
$(X_{k,\hspace*{1mm} a})$ for each sub-period $I_{a}$, defined as
\begin{equation}
X_{k,\hspace*{1mm}a} = \sum_{i=1}^{k}(N_{i,\hspace*{1mm} a}
-M_{a}), \hspace*{1cm}k = 1, 2, ...n.
\end{equation}
As can be seen from Eq. (3), the series of accumulated departures
from the mean always will end up with zero. Now, the range that
the time series covers relative to the mean within each sub-period
is defined as
\begin{equation}
R_{I_{a}} = \max(X_{k,\hspace*{1mm}a}) -
\min(X_{k,\hspace*{1mm}a}),\hspace*{1cm} 1 < k < n.
\end{equation}
The next step is to calculate the standard deviation for each
sub-period $I_{a}$,
\begin{equation}
S_{I_{a}}
=\sqrt{\frac{1}{n}\sum_{k=1}^{n}(N_{k,\hspace*{1mm}a}-M^{^{^{2}}}_{a}
)}.
\end{equation}
Then, the range for each sub-period $(R_{I_{a}})$ is rescaled by
the corresponding standard deviation $(S_{I_{a}})$. Recall that we
had $m$ contiguous sub-periods of length $n$. Thus, the average
$R/S$ value for length or box $n$ is
\begin{equation}
(R/S)_{n}= \frac{1}{m}\sum_{a=1}^{m}( \frac{R_{I_{a}}}{S_{I_{a}}}
).
\end{equation}
Now, the calculations from Eqs. (1)-(6) must be repeated for
different time horizons. This is achieved by successively
increasing $n$ and repeating the calculations until we have
covered all integer $n's$. One can say that $R/S$ analysis is a
special form of \emph{box-counting} for time series. However, the
method was developed long before the concepts of fractals. After
having calculated $R/S$ values for a large range of different time
horizons $n$, we plot $\log(R/S)_{n}$ against $\log(n)$. By
performing a least-squares regression with $\log(R/S)_{n}$ as the
dependent variable and $\log(n)$ as the independent one, we find
the slope of the regression which is the estimate of the
\emph{Hurst exponent} $H$. The Hurst exponent ($H$) and the
fractal dimension $D_{f}$ are related as \cite{Peters}
\begin{equation}
D_{f} =2 - H.
\end{equation}
In theory, $H=0.5$ means that the time series is independent, but
as mentioned above the process need not be Gaussian. If $H=0.5$,
the process may in fact be a non-Gaussian process as e.g. the
Student-t or gama. If $H \in (0.5, 1.0]$ it implies that the time
series is persistent which is characterized by long memory effects
on all time scales. For example, all daily price changes are
correlated with future daily price changes; all weekly price
changes are correlated with all future weekly price changes and so
on. This is one of the key characteristics of fractal time series.
It is also a main characteristic of non-linear dynamical systems
that there is a sensitivity to initial conditions which implies
that such a system in theory would have an infinite memory. The
persistence implies that if the series has been up or down in the
last period then the chances are that it will continue to be up or
down, respectively, in the next period. This behavior is also
independent of the time scale we are looking at. The strength of
the trend-reinforcing behavior, or persistence, increases as $H$
approaches 1.0. This impact of the present on the future can be
expressed as a correlation function ($C$),
\begin{equation}
C = 2^{(2H-1)} - 1.
\end{equation}
In the case of $H=0.5$ the correlation $C$ equals zero, and the
time series is uncorrelated. However, if $H=1.0$ we see that
$C=1$, indicating perfect positive correlation. On the other hand,
when $H \in [0, 0.5)$ we have anti-persistence. This means that
whenever the time series have been up in the last period, it is
more likely that it will be down in the next period. Thus, an
anti-persistent time series will be more choppier than a pure
random walk with $H=0.5$. The $R/S$ analysis can also uncover
average non-periodic cycles in the system under study. If there is
a long memory process at work, for a natural system this memory is
often finite, even though  long memory processes theoretically are
supposed to last forever, as was the case for mathematical
fractals and the logistic map. When the long term memory is lost,
or the memory of the initial conditions has vanished, the system
begins to follow a random walk; this is also called the crossover
point. Thus, a crucial point in the estimation of the Hurst
exponent is to use the proper range for which there is non-normal
scaling behavior. This is the range for which the scaling behavior
is \emph{linear} in the $\log(R/S)_{n}$ versus $\log(n)$ plot. If
there is a \emph{crossover-point}, this can be seen as a
\emph{break} in the plot where the slope changes for a certain
value, $\log(n_{max})$. If this is the case, it is an indication
of a non-periodic cycle with average cycle length equal
to $n_{max}$.\\

A plot of the rescaled range $R/S$ as a function of $\tau$ for the
TEPIX returns over the mentioned period is shown in the curve of
Fig.2. The data in this case show a scaling regime that goes from
$\tau=2$ up to 7 (in $\log$ scale) approximately. It is equal to
128 trading days or 180 days. A linear regression in this region
yields the value $H=0.79\pm0.03$. It must be mentioned that the
Hurst method tends to overestimate the Hurst exponent for time
series of small sizes \cite{J. Feder}.

\begin{figure}[ht]
{\centering
\resizebox*{0.7\textwidth}{0.35\textheight}{\includegraphics{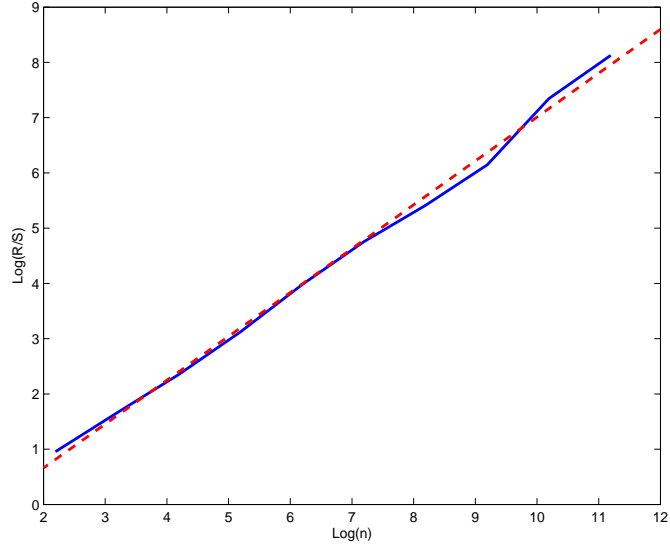}}
\par} \caption{Rescaled range $R/S$ versus the time lag $\tau$ for the
returns of the TEPIX in the period of 1995-2004.}
\end{figure}

\subsection{Lo's Modified $R/S$ Analysis}
The classical $R/S$ test has been proven to be too weak to
indicate a true long memory process, in fact it tends to indicate
a time series has long memory when it does not. In 1991 Lo
\cite{Lo} introduced a stronger test based on a modified $R/S$
statistics, which is known to be too strong to indicate a true
long memory
process. Lo's modified $R/S$ test is described in brief below.\\

For a given series $N_{i}, i=1,2,...n$, Lo defines the modified
$R/S$ statistic as,
\begin{equation}
Q_{n}=\frac{[\max\sum_{j=1}^{k}(N_{j}-\bar{N_{n}}\hspace{1mm}
)-\min\sum_{j=1}^{k}(N_{j}-\bar{N_{n}}
\hspace{1mm})]}{\sigma_{n}(q)},\hspace*{10mm} k=1,n
\end{equation}
where the denominator is expressed as,
\begin{equation}
\sigma^{2}_{n}(q)=\frac{1}{n}\sum_{j=1}^{k}(N_{j}-\bar{N_{n}}\hspace{1mm}
)^{2}+\frac{2}{n}\sum_{j=1}^{q}w_{j}(q)[\sum_{i=j+1}^{n}(N_{i}-\bar{N_{n}}\hspace{1mm}
)(N_{i-j}-\bar{N_{n}}\hspace{1mm} )]
\end{equation},
in which
\begin{equation}
w_{j}(q)=1-\frac{j}{q+1}, \hspace*{10mm}q<n
\end{equation}
Lo finally standardizes the statistic $Q_{n}$ by dividing by
$\sqrt{n}$ and is denoted as $V_{n}(q)$. The numerator of
$V_{n}(q)$ is the range of deviation from the approximate linear
trend line in a given interval and the denominator is the sample
variance augmented with weighted autocovariances up to a lag
determined $q$. For $q=0$, this is same as the classical $R/S$
statistic. This autocovariance part of the denominator is non zero
for series exhibiting short term
memory and this makes the statistic robust to heteroscedasticity.\\

Table 2 gives the results from the modified $R/S$ statistic. $R/S$
analysis is extremely sensitive to the order of truncation $q$ and
there is no statistical criteria for choosing $q$ in the framework
of this statistic. Since there is no data driven guidance for the
choice of this parameter, we consider different values for
$q=0,2,4,6,8,10$ and $15$. More explanations related to long
memory effect and interpretation of data in Table 2 will be
mentioned in Section 7. In brief, the starred values in Table 2
reject null hypothesis of short memory.

\begin{table}[htb]
\begin{center}
\caption{\label{Tb2}Modified rescaled range ($R/S$) statistic for
the returns, absolute and squared returns.}
\medskip
\begin{tabular}{|c|ccc|}

\hline\ Lag order&$R/S$ statistic&&\\\hline $q$&returns&absolute
returns&squared returns\\\hline
$0$&$1.0513$&$0.7857^{\ast}$&$0.9573$\\\hline
$2$&$0.8219$&$0.5393^{\ast}$&$0.7528^{\ast}$\\\hline
$4$&$0.7475^{\ast}$&$0.4475^{\ast}$&$0.6654^{\ast}$\\\hline
$6$&$0.7156^{\ast}$&$0.3952^{\ast}$&$0.6111^{\ast}$\\\hline
$8$&$0.6964^{\ast}$&$0.3608^{\ast}$&$0.5734^{\ast}$\\\hline
$10$&$0.6807^{\ast}$&$0.3362^{\ast}$&$0.5451^{\ast}$\\\hline
$15$&$0.6724^{\ast}$&$0.2972^{\ast}$&$0.4980^{\ast}$\\\hline

\end{tabular}
\end{center}
\end{table}

\section{Detrended Fluctuation Analysis}
Detrended fluctuation analysis (DFA) is a scaling analysis
technique providing a simple quantitative parameter-the scaling
exponent $\alpha$-to represent the correlation properties of a
time series \cite{C.K. Peng}. The advantage of DFA over many
techniques are that it permits the detection of long-range
correlations embedded in seemingly non-stationary time series, and
also avoids the spurious detection of apparent long-range
correlations that are an artifact of non-stationarity.
Additionally, the advantages of DFA in computation of $H$ over
other techniques (for example, the Fourier
transform) are:\\
\begin{itemize}
\item inherent trends are avoided at all time scales; \item local
correlations can be easily probed.
\end{itemize}

To implement the DFA, let us suppose we have a time series, $N(i)
(i=1,...,N_{max})$. We integrate the time series $N(i)$:
\begin{equation}
y(j)=\sum_{i=1}^{j}[N(i)-\langle N \rangle]
\end{equation}
where:
\begin{equation}
\langle N\rangle=\frac{1}{N_{max}}\sum_{j=1}^{N_{max}}N(i).
\end{equation}
Next we break up $N(i)$ into $K$ non-overlapping time intervals,
$I_{n}$, of equal size $\tau$ where $n=0,1,...K-1$ and $K$
corresponds to the integer part of $N_{max}/\tau$. In each box, we
fit the integrated time series by using a polynomial function,
$y_{pol}(i)$, which is called the local trend. For order-\emph{l}
DFA (DFA-1 if $l$=1, DFA-2 if $l$=2, etc.), the \emph{l}-order
polynomial function should be applied for the fitting. We detrend
the integrated time series $y(i)$ in each box, and calculate the
detrended fluctuation function:
\begin{equation}
Y(i)=y(i)-y_{pol}(i).
\end{equation}
For a given box size $s$, we calculate the root mean square
fluctuation:
\begin{equation}
F(s)=\sqrt{\frac{1}{N_{max}}\sum_{i=1}^{N_{max}}[Y(i)]^{2}}
\end{equation}
The above computation is repeated for box sizes $s$ (different
scales) to provide a relationship between $F(s)$ and $s$. A power
law relation between $F(s)$ and $s$ indicates the presence of
scaling: $F(s)\sim s^{\alpha}$. The parameter $\alpha$, called the
scaling exponent or correlation exponent, represents the
correlation properties of the signal: if $\alpha=0.5$, there is no
correlation and the signal is an uncorrelated signal \cite{C.K.
Peng}; if $\alpha<0.5$, the signal is anticorrelated; if
$\alpha>0.5$, there are positive correlations in the signal. In
the two latest cases, the signal can be well approximated by the
fractional Brownian motion law \cite{J. Feder}.\\

In Fig.3 we plot in double-logarithmic scale the corresponding
fluctuation function $F(s)$ against the box size s. Using the
above procedure, we obtain the following estimate for the Hurst
exponent: $H=0.72\pm 0.01$. Since $H>0.5$ it is concluded that the
TEPIX returns show persistence; i.e, strong correlations between
consecutive increments. It is seen that for $s\sim115$ the
empirical data deviate from the initial scaling behavior. This
indicates that the TEPIX tends to loose its \emph{memory} after a
period of about 162 days. Based on overestimating the Hurst
exponent in the $R/S$ analysis it may be explained why the
exponent $H$ obtained via the Hurst method is usually larger than
that of the DFA method \cite{P. Grau}.

\begin{figure}[htb]
{\centering
\resizebox*{0.7\textwidth}{0.35\textheight}{\includegraphics*{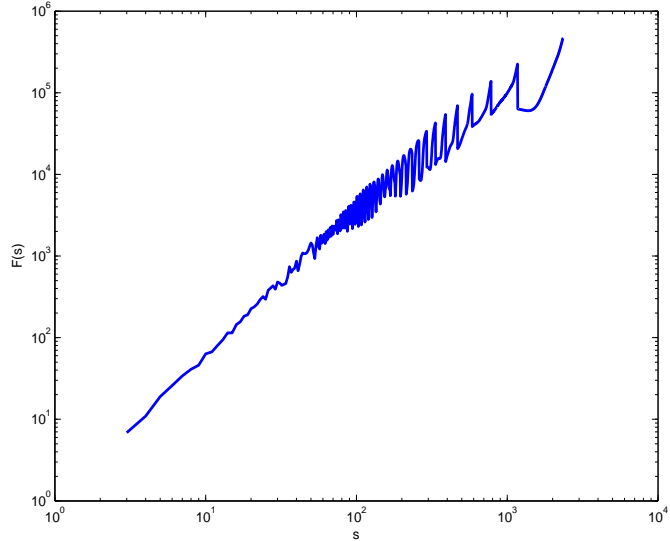}}
\par} \caption{Fluctuation Function $F(s)$ as a function of box size
for the returns of the TEPIX index in the period of 1994-2004.}
\end{figure}

\section{Generalized Hurst Exponents Approach}
A generalization of the approach proposed by Hurst should be
associated with the scaling behavior of statistically significant
variables constructed from the time series \cite{Barabasi1}.
Therefore we analyze the $q$-order moments of the distribution of
the increments which is a good characterization of the statistical
evolution of a stochastic variable $P(t)$. The generalized Hurst
exponents, $H_{q}\equiv H(q)$, for a time series $P(t)
(t=1,2,...)$ are defined by the scaling properties of its
structure functions $S_{q}(\tau)$
\begin{equation}
S_{q}(\tau) =
\langle|P(t+\tau)-P(t)|^{q}\rangle_{T}^{\frac{1}{q}}\sim
\tau^{H(q)}
\end{equation}
where $q > 0$, $\tau$ is the time lag and averaging is over the
time box (window) $T \gg\tau$, usually the largest time scale of
the system. The function $H(q)$ contains information about
averaged generalized volatilities at scale $\tau$ (only $q=1,2$
are used to define the volatility). In particular, the $H(1)$
exponent indicates persistent ($H(1) > 0.5$) or anti-persistent
($H(1) < 0.5$) behavior of the trend. For the Brownian random walk
one gets $H(1)= 0.5$. For the popular L\'{e}vy stable and
truncated L\'{e}vy processes with parameter $\alpha$, it has been
found that $H(q) = q/\alpha $ for $q < \alpha $ and $H(q) = 1$ for
$q \geq\alpha$. In this framework, we can distinguish between two
kinds of processes:
\begin{itemize}
\item a process where $H(q)=H$, constant independent of $q$; \item
a process with $H(q)$ not constant.
\end{itemize}
The first case is characteristic of unifractal processes where
$qH(q)$ is linear and completely determined by its index $H$. In
the second case, when $H(q)$ depends on $q$, the process is
commonly called multi-fractal and different exponents characterize
the scaling of different $q$-moments of the distribution.\\

Eq. (13) is studied numerically in order to analyze the
generalized $q$th-order Hurst exponents in the structure function
$S_{q}(\tau)$. Table 3 includes the values of the generalized
Hurst exponents $H(q)$ in the structure function for the TEPIX.
The values $H(q)$ versus $q$ for $q=1,...,10$ are plotted in Fig.
3 for the TEPIX.

\begin{figure}[htb]
{\centering
\resizebox*{0.7\textwidth}{0.35\textheight}{\includegraphics{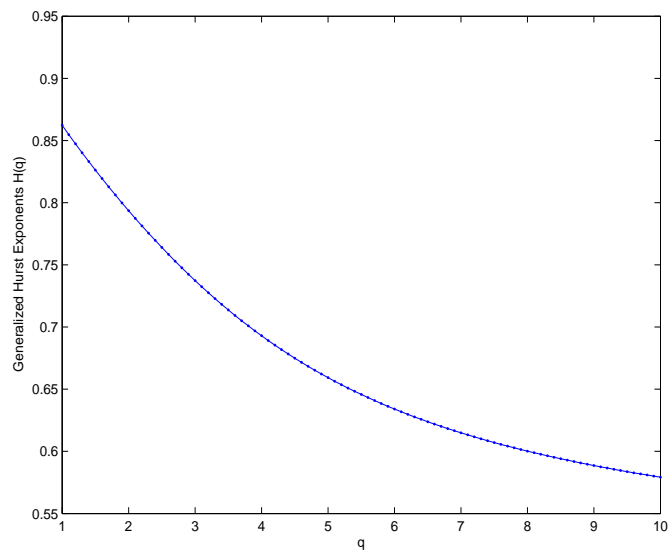}}
\par} \caption{Generalized Hurst exponents versus the $q$ for the returns
of the TEPIX in the period of 1995-2004.}
\end{figure}

\begin{table}[htb]
\begin{center}
\caption{\label{Tb2}Values of the generalized $q$th-order Hurst
exponents $H(q)$ for the TEPIX.}
\medskip
\begin{tabular}{|c|c|c|c|c|}
\hline $H(1)$&$H(2)$&$H(3)$&$H(4)$&$H(5)$\\\hline
0.8622&0.7935&0.7373&0.6929&0.6592\\\hline\hline
$H(6)$&$H(7)$&$H(8)$&$H(9)$&$H(10)$\\\hline
0.6340&0.6149&0.6001&0.5885&0.5792\\
\hline

\end{tabular}
\end{center}
\end{table}

It should be noted that the methods listed above, that is, $R/S$
analysis, Lo's $R/S$ analysis and DFA can only extract a single
scaling exponent from a time series. However, it is possible that
the given time series may be governed by more than one scaling
exponents, in which case a single scaling exponent would be unable
to capture the complex dynamics inherent in the data. Analysis
using generalized Hurst exponents method, elucidates the
dependence of $H(q)$ on $q$, which is a hall mark of multifractal
processes. Such processes are far more than one exponent to
characterize their scaling properties \cite{Ivanov}.

\section{Characteristic Exponent}

Paul L\'{e}vy, the French mathematician, proposed a general
approach with the Gaussian as only a special case, to identify
probability distributions which their sum has the same probability
distribution. A stable L\'{e}vy distribution is represented by
\cite{Mantegna}
\begin{equation}
L_{\alpha}(N, \triangle t)\equiv\frac{1}{\pi}\int_{0}^{\infty}
\exp(-\gamma\triangle t q^{\alpha})\cos(qN)dq
\end{equation}
where $\alpha$ is the characteristic exponent $0<\alpha\leq 2$, $
N$ the return, $\gamma$ the scale factor, and $\triangle t$ the
time interval.\\
This distribution obeys below scaling relations:
\begin{equation}
N_{\triangle t}=N_{s}(\triangle t)^{1/\alpha}
\end{equation}
and
\begin{equation}
L_{\alpha}(N_{s},\triangle t)\equiv L_{\alpha}(N_{s},1)(\triangle
t)^{-1/\alpha}.
\end{equation}
If $\alpha=2$ the distribution is Gaussian, and there is a finite
second moment. If $\alpha=1$ we have the Cauchy distribution with
both infinite first and second moments. In the region for which
$1<\alpha< 2$, the second moment becomes infinite, but with a
stable mean. L\'{e}vy stable distributions are self-similar and
this means that the probabilities of return are the same for all
time intervals once we adjust for the time scale. Roughly
speaking, an agent with 1 min time interval faces the same risk as
a 100 min agent in his time interval when adjusted for scale.
The $\alpha$ exponent takes this scaling relationship into account.\\
The fractal dimension of the probability space, $\alpha$, used in
above Equations is related to the Hurst exponent of the time
series as:
\begin{equation}
\alpha=\frac{1}{H}.
\end{equation}

In this way, characteristic exponent for return distribution of
TEPIX can be calculated. $\alpha$ exponents derived by using all
of above methods has been shown in Table 4.

\begin{table}[htb]
\begin{center}
\caption{\label{Tb2}Values of the $\alpha$ exponents resulted from
Hurst exponents. }
\medskip
\begin{tabular}{cccc}
\hline\hline $\alpha_{ R/S}$&$\alpha_{Lo}$&$\alpha_{
DFA}$&$\alpha_{ GHE}$\\\hline
1.266&1.389&1.163&1.160\\
\hline\hline
\end{tabular}
\end{center}
\end{table}

These estimates of $\alpha$ are relatively close to each other.
Based on $\alpha$ values, we observe a non-normal scaling behavior
and all estimates indicate that the process is different from a
pure random walk. In fact, these results are evidence of a
non-linear chaotic system.\\
The distribution of TEPIX returns can be fitted by a stable
L\'{e}vy distribution. For a better comparison of the return
distribution with a Gaussian PDF and evaluating derived Hurst
exponents, we have performed a maximum likelihood estimation of
stable parameters. The parameters of this fitted L\'{e}vy
distribution is presented in Table 5.\\

Alpha is the same characteristic exponent. Beta is the skewness in
the range [-1,1] and gamma and delta are straightforward scale and
shift parameters respectively. A probability distribution function
of returns against Gaussian distribution with the mean and
standard deviation of fitted L\'{e}vy distribution is depicted in
Fig.5. Fitted L\'{e}vy distribution iself, is plotted in the
Fig.6. It can be seen from this figures that the the real
distribution (or the L\'{e}vy fitted ones) of
returns is different from a Gaussian PDF (random walk).\\

\begin{figure}[htb]
{\centering
\resizebox*{0.7\textwidth}{0.35\textheight}{\includegraphics{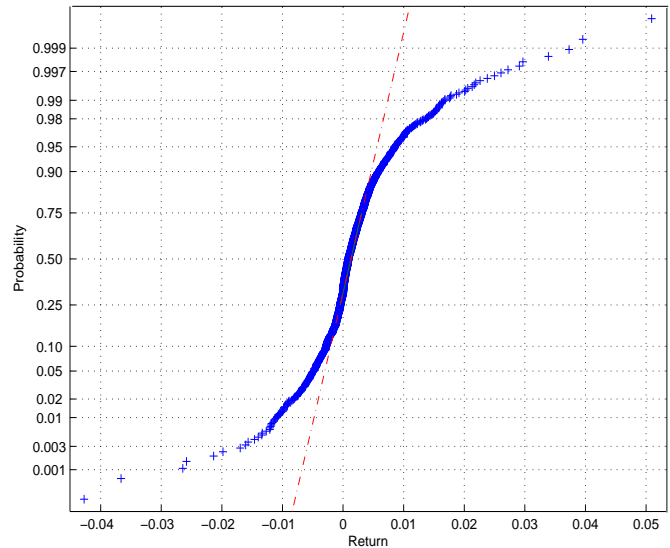}}
\par} \caption{Probability distribution function of returns against a Gaussian distribution .}
\end{figure}

\begin{figure}[htb]
{\centering
\resizebox*{0.7\textwidth}{0.35\textheight}{\includegraphics{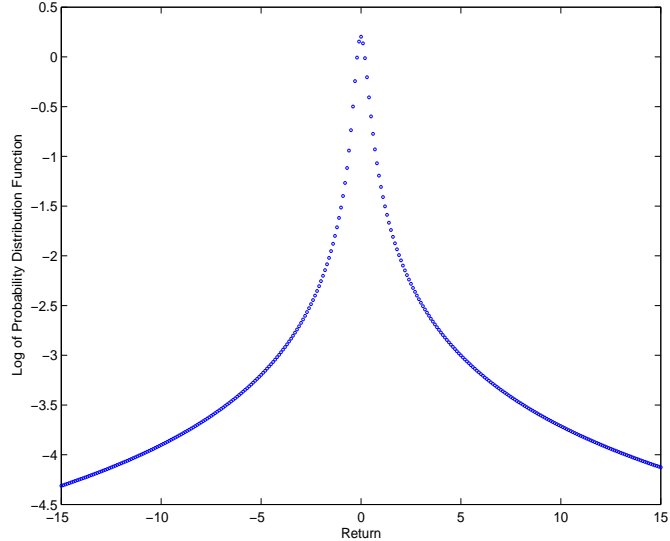}}
\par} \caption{Probability Distribution Function(PDF) for daily returns
of TEPIX (vertical axis is in log scale).}
\end{figure}
\begin{table}[htb]
\begin{center}
\caption{\label{Tb2}The parameters of the fitted L\'{e}vy
distribution.}

\begin{tabular}{ccccccccccccc}
\hline\hline $\alpha$&$\beta$&$\gamma$&$\delta$\\\hline
1.316 & 0.2049 & 0.0018 & 0.00076 \\
\hline\hline
\end{tabular}
\end{center}
\end{table}

Based on maximum likelihood parameter estimation of L\'{e}vy
distribution (direct estimation), the best characteristic exponent
has been resulted by classical $R/S$ analysis and Lo's $R/S$
analysis respectively.

\section{Long Memory Process}

A random process is called a long memory process if it has an
autocorrelation function that is not integrable. This happens, for
example, when the autocorrelation function decays asymptotically
as a power law of the form $\tau^{-\alpha}$ with $\alpha < 1$.
This is important because it implies that values from the distant
past can have a significant effect on the present, and implies
anomalous diffusion. A process is defined as long memory if in the
limit $k\rightarrow\infty$
\begin{equation}
\gamma(k)=\sim k^{-\alpha}L(k)
\end{equation}

where $ 0 < \alpha < 1$ and $L(x)$ is a slowly varying function at
infinity. The degree of long memory dependence is given by the
exponent $\alpha$; the smaller $\alpha$, the longer the memory.\\
The Hurst exponent simply is related to $\alpha$. For a long
memory process $H = 1 - \frac{\alpha}{2}$ or $\alpha = 2 - 2H$.
Short memory processes have $H = \frac{1}{2}$, and the
autocorrelation function decays faster than $k^{-1}$. A positively
correlated long memory process is characterized by a Hurst
exponent in the interval (0.5, 1).\\
As an application of results provided by Lo's modified $R/S$
statistic, it can be said \cite{Lo2}, at the 5\% significance
level, the null hypothesis of a short memory process is rejected
if the modified $R/S$ statistic does not fall within the
confidence interval [0.809, 1.862]. For returns, the null
hypothesis of short memory is rejected at any lags, except for 0
and 2. For absolute and squared returns, the null hypothesis of
short memory is rejected for all
lag orders.\\
Besides, the Hurst exponent is linked to the modified $R/S$
statistic by
\begin{equation}
\lim_{T\rightarrow +\infty}E[R_{T}/S_{T}(q)]/(aT^{H})=1
\end{equation}
with $a>0$. With this link it is possible to obtain the following
approximate relationship:
\begin{equation}
\log{E[R_{T}/S_{T}(q)]}\cong \log(a)+H \log(t)
\end{equation}
In order to estimate the value of the Hurst exponent, $H$, we
first determine a series of estimates of the Hurst exponent by
fitting an ordinary least squares regression between
${\log[R_{T,l}/S_{T,l}(q)],l=1,...,j}$ and ${\log(l),l=1,...,j}$
for every $j=2,...,T^{\ast}$, where $R_{T,l}$ and $S_{T,l}(q)$ are
quantities related to $R_{T}$ and $S_{T}(q)$, respectively. Then
we choose the optimal estimate in this series. As a result of such
procedure, Hurst exponent has obtained equal to 0.721$\pm$0.001.
Therefore, the Lo's method verifies long memory process in returns
of the TEPIX, based on above discussions.

\section{Market Stage of Development}

Based on recent research for characterizing the stage of
development of markets \cite{T.Di Matteo} it is shown that the
exponent $H(2)$ has sensitivity to the degree of development of
the market. As far as Stock markets are concerned, the generalized
Hurst exponents $H(1)$, $H(2)$ show remarkable differences between
developed and emerging markets. At one end of the spectrum there
are stocks like as the Nasdaq 100 (US), the S\&P 500 (US), the
Nikkei 225 (Japan) and so on. Whereas , at the other end, there
are Russian AK\&M, the Indonesian JSXC, the Peruvian LSEG, etc.
This sensitivity of the scaling exponents to the market conditions
provides a new and simple way of empirically
characterizing the development of financial markets.\\
Roughly speaking, emerging markets are associated with high value
of $H(1)$ and developed markets are associated with low values of
$H(1)$. Besides, it is found that all the emerging markets have
$H(2)\geq 0.5$ whereas all the developed have $H(2)\leq 0.5$.\\
Considering all of above discussions and results, we notice that
Tehran Stock Exchange belongs to emerging markets category and it
is far from an efficient and developed market. Hurst exponent
calculated by $R/S$ and DFA methods in one hand and generalized
Hurst exponents ($H(1)$ and $H(2)$) in the other hand, both
present
this fact.\\
For the sake of comparison between various stock markets, the two
first generalized Hurst exponents are shown in Table 5. It must be
noticed that all of data in Table 6, except those corresponds to
TEPIX, have been calculated during 1997 to 2001 period \cite{T.Di
Matteo}, while those corresponds to TEPIX have been calculated in
the time period from 1995 to 2004.

\begin{table}[htb]
\begin{center}
\caption{\label{Tb2}Hurst exponents $H(1)$ and $H(2)$ for stock
market indices.}
\medskip
\begin{tabular}{c|c|c}

\hline\hline

$$Stock $ $Market $ $indices$$&$H(1)$&$H(2)$\\\hline
Nasdaq 100&0.47&0.45\\
S\&P 500&0.47&0.44\\
Nikkei 225&0.46&0.43\\
AK\&M&0.65&0.51\\
JSXC&0.57&0.53\\
LSEG&0.61&0.58\\
TEPIX&0.86&0.79\\
\hline\hline
\end{tabular}
\end{center}
\end{table}

These results indicate that, financial market characteristics in
Iran do not show developed situations. In fact, Tehran Stock
Exchange belongs to the category of emerging financial markets.

\section{Conclusions}
In this paper the concept of multifractality has been applied to
Tehran Stock Exchange data. This market show a fractal scaling
behavior significantly different from what a random walk would
produce. For TEPIX time series we have obtained a Hurst exponent
greater than 0.5, indicating that the TEPIX has long term
dependence (persistence). This memory effect seems to last for up
to about 6 months (115-128 trading days). Analysis using
generalized Hurst exponents method, indicates the dependence of
$H(q)$ on $q$, which is an evidence of multifractal processes.
Also, we show that based on generalized Hurst exponents, financial
market characteristics in Iran do not indicate a developed market.
In other words, we are dealing with an emerging capital market.
These findings imply that there are patterns, or trends in returns
that persist over time. This provides a theoretical platform
supporting the use of technical analysis to produce above average
returns. The findings may be used to improve the current models or
to make new ones which use the concept of fractal scaling.

\section{Acknowledgment}
We thank A.T. Rezakhani for reading manuscript and giving helpful
comments.

\end{document}